# Datastringer: easy dataset monitoring for journalists

Matt Shearer
BBC News Labs
One Euston Square
40 Melton St, London, NW1 7TQ
matt.shearer@bbc.co.uk

Basile Simon
BBC News Labs
One Euston Square
40 Melton St, London, NW1 7TQ
basile.simon01@bbc.co.uk

Clément Geiger
BBC News Labs
176 rue Philippe de Commynes
45160 Olivet, France
clement.geiger@gmail.com

## ABSTRACT

We created a software enabling journalists to define a set of criteria they would like to see applied regularly to a constantly-updated dataset, sending them an alert when these criteria are met, thus signaling them that there may be a story to write. The main challenges were to keep the product scalable and powerful, while making sure that it could be used by journalists who would not possess all the technical knowledge to exploit it fully. In order to do so, we had to choose Javascript as our main language, as well as designing the code in such a way that it would allow re-usability and further improvements. This project is a proof of concept being tested in a real-life environment, and will be developed towards more and more accessibility.

## General Terms

Algorithms

## 1. INTRODUCTION

In our ongoing research about the future of reporting and how to make our journalists' every day work easier, we noticed a lack of tool to monitor the information coming in the form of datasets.

Data-journalism now has its own place in the mainstream media newsrooms, but working with data is still reserved to data-journalists. This is largely caused by the nature of this source, that not every journalist knows how to handle properly.

We wanted to create a computer-assisted reporting tool that would simplify the process of using data as a source, and to make that process as simple as possible while remaining powerful. We created Datastringer, a tool which, once configured, sends alerts to journalists when pre-defined events appear in the data, while remaining invisible the rest of the time.

Several tools already existed to work in such a way. We aimed to innovate by creating our own with simplicity and usability by all users in mind. Our goal is to make the tool more widely adopted across the BBC and in the media in general, by providing a lightweight and hopefully rather straightforward solution to the problem of surfacing the interesting news in a constantly updated dataset.

Datastringer and its first tests hope to prove that this tool can be used effectively by journalists to get original stories with a minimal effort of research and pre-definition of the story they are looking for.

## 2. MONITORING DATA SOURCES

Addressing the new streaming nature of data was one of the main goals of the New York Times Labs' "Stream Tools." [1] We also came to realise that the publication of large datasets under the Open Government Licence didn't mean that journalists were making an optimal use of these sources.

Consuming this data is inherently harder than reading a block of text appearing on a AP news-ticker. It requires a certain education to numbers, and the datasets can be quite big, complex, even obscure at time.

More problematic, it is even harder to discernate updates in a dataset. The numbers changed between the two versions, but the two sets (the new one and the old one) are not always available at the same time, and their large size can be a hurdle to an easy comparison, one that would be done with a naked eye.

We aimed to design Datastringer in a way that would make the comarison of datasets and the changes in the data the main appeal to journalists, so the news they would report would not be the dataset's publication itself, but the new data contained in this dataset, or what these new figures mean.

This step of downloading the data, inspecting it and comparing it, is often very repetitive, and quite imprecise if the journalist doesn't know what to look for. PR releases already provide bullet-points and summaries about the interesting bits of it, and journalists could find themselves in the position where, each time a new set is released, they have to do certain calculations with it to see if things changed, and if yes, by how much.

A programmatic solution was then required to automate the process. The solution we explored is to define once the calculations to operate on the dataset, to automate the updated dataset query and the above calculations, then to trigger a system of alert/notification.

## 3. MAKING IT EASY
### 3.1. Product philosophy

Other solutions existed to consume data in such a way. OpenNews Fellow Annabel Church's Datawi.re [2], in particular, was built with exactly the same desire in mind – in fact, our idea was greatly inspired by a Church's workshop we attended. However, Datawi.re has a relatively high barrier of entry: it requires PostgreSQL, AWS, Twitter and Facebook apps... We intended Datastringer to be used directly by the journalists, and so we needed to adapt to their technical knowledge and resources.

Simplicity was, in our minds, absolutely paramount. It also meant designing a workflow that will be completely transparent for the user, and which could be summarised by: "configure it once, forget about it", as the user delegated the repetitive tasks to the machine.

## 3.2. Choosing Javascript

The choice of the language was an important decision. This language would eventually have to be used, to a certain extent, by the end-user, and we wanted to avoid assuming high technical skills.

The choice of Javascript and Node.js was the only logical one, as it represented a compromise between availability, simplicity, and common knowledge. Interacting with basics layers of Javascript is quite difficult to avoid for online journalists who have to deal with embedded contents, widgets, and other small snippets of code found online. Hopefully, these journalists will be familiar enough with the language or its syntax to understand what's happening. We will see later in this paper what can be done for those who do not have this knowledge at all.

## 3.3. Choosing the email for alerts

For journalists un-familiar with this kind of monitoring, email as an alert medium was a logical choice. It provides a familiar environment with which the journalist interacts, and keeps the leg-work completely un-obstrusive.

The use of UNIX Postfix [3] and Node.js' Nodemailer [4] module allows us to send emails from the client's machine, without having to make the user's address transit online – thus avoiding security risks.

## 3.4. Cron scheduling for automation

Un-obstrusiveness was accomplished by automating Datastringer's execution, i.e. by shceduling it to run regularly without requiring any kind of user's intervention.

We used UNIX's popular solution Cron [5] to add to the user's configuration a job scheduled every day at 12:00, directly from the installation.

## 4. SCALABILITY AND POWER-USAGE

We wanted to keep Datastringer simple while designing it in a way that would satisfy power-users and hacks more proficient in Javascript and computer-assisted reporting. For the product to appeal to several categories of users, it was necessary to have large layers of customisation available.

## 4.1. File structure

The main program, `datastringer.js`, should be considered as a "black box" using user-inputted values, parameters, and functions, stored in `use_cases.json`. A look at this file's structure (see illustration 1) provides at a glance a relatively good idea of what the program is going to execute. It first links to another Javascript file, which contains the main methods to query the data and perform calculations on it, as well as defining the parameters, which are passed in this file in the following line.

```
[{
  "stringer": "local-police-stringer.js",
  "parameters": ["metropolitan", "00AGGU"]
}, {
  "stringer": "crime-stringer.js",
  "parameters": ["51.52863195218981", "-0.12342453002929688", "6", "10"]
}]
```

**Illustration 1: use_cases.json structure**

As mentioned earlier, the stringer defines the core function attached to the data. Illustration 2 shows an example of how such a function can be defined.

```
function stringer(lat, lng, numberOfMonths, threshold, callback) {
  // get the data for each month
  // sort crimes in categories
  // for each category, calculate the average on y months, y being numberOfMonths
  // compare the last month value and the average
  // if this difference is more than x % or less than -x %, trigger the alert, x being threshold
}
```

**Illustration 2: a stringer structure**

## 4.2. Stringer creation

If separating Datastringer's workload into two files is not obvious, the file structure designed this way provides a lot of freedom to the user, and promotes scalability.

Should the user want to create his own stringer, he would have to write his own function, similar or different to the example shown in Illustration 2. By leaving this definition work to the user, the programs offers all of Javascript power and functionalities to Datastringer. Thus, talking to a REST JSON API, an open-government triplestore returning CSV files, or an RSS feed, are only part of what can be done with Datastringer.

By defining the stringer function, the user also controls its level of complexity. One of the examples provided with Datastringer downloads a JSON file, compares it with the previous version stored locally, and returns the differences between the two for an alert. The second one, way more complex, will query a JSON file listing all the crimes in a neighbourood, then aggregate the results by sorting and ranking the categories of crime, then compare this month's numbers to an average calculated by querying and performing the same operation on the previous $x$ months, and finally check if this difference is greater than $-y$ % - where $x$ and $y$ are defined by the user).

## 4.3. Re-usability

The user also has the option to define as many parameters to replicate his stringer. Nothing prevents him from hard-coding these parameters to his stringer function, then not to pass parameters in the `use_cases` file.

However, if parameters are defined, then the stringer becomes completely re-usable. The two examples we provide require at least GPS coordinates to monitor a neighbourhood. Because these two informations (`lat` for latitude, `lon` for longitude) are parametrised, the user can monitor as many neighbourhoods as he wants, just by copy-pasting in his `use_cases` file the correct parameters: an array calling the stringer javascript file and the new parameters, on the model of Illustration 1.

## 5. IMPROVEMENTS

As it is today, Datastringer is a proof of concept. It hopes to demonstrate the viability of this data consumption process and the added-value of monitoring datasets. However, many improvements are required to deliver all the promises the program makes.

The simplicity announced in our philosophy is quite not there yet. The installation still involves manipulating the command line and using Git – a knowledge we cannot possibly expect from every journalist. It should be made simpler and more straightforward, by the familiar download of a ZIP archive and installation by double-click.

We would also like to unshackle working with Datastringer from any sort of coding, should the user be willing to do so. The solution to find here is to build a graphical user-interface (GUI), that will have several functionalities:

1. The management of the stringers and use cases. The user should be able to turn on and off a stringer and its set of parameters, via a simple button. Duplicating a stringer to assign it different parameters should also be possible. This dashboard-style idea should be quite simple to design, given that the page's only goal is to write and edit a JSON file.
2. The creation of stringers and use cases. The user should be invited to fill text-boxes with informations about the considered data source, then click on the relevant figure. This would be more complex to implement, but we would like to get inspiration from the process followed by an Import.io user to create web-scrapers graphically.

The program should also work on several, if not all platforms. This cross-platform development is already a work in progress, and involves large changes in the code-base and in the chosen tools. As an example, cron jobs are not available on Windows, and would need to be replace by an equivalent solution for scheduling jobs.

## 6. CONCLUSION

Datastringer received a lot of interest during its pre-launch, both on Hacker News (rising to the third link on the front-page for several hours) and on Github (being ranked in the Trending category for a day). We presented the project to some of our broadcast and local journalists, who all expressed interest for the tool and for what they believed could be of great help to their research work, by making the surfacing of stories automatic.

We are currently assisting a team of journalists from BBC London local station in the creation of their stringers, and the monitoring of crime statistics for the whole Greater London. The idea for their stringer being that each alert received by them must be a good story to write, and contain the headline elements (ex: "Bicycle theft on the rise by 34% in London").

Another lead we are considering is to publish a "software-as-a-service" [6] version, centrally hosted and accessed by the user via a graphical browser interface only.

We are also working on more complex examples involving mash-ups of datasets, proving once again Datastringer's flexibility and hackability for power-users.

The biggest limit of Datastringer's functionality is that the user needs to pre-define what he is looking for in the data. This effort of original research can be quite daunting to those unfamiliar with dataset research and work, and will probably reserve its use to those keen on this practice. Also, the pre-definition of the parameters does not leave any place to the discovery of an unexpected and original story. At the time, Datastringer is only a monitoring tool, and the users must be aware that it will only surface what they are looking for.